\documentclass{PoS}

\title{New results on the search for spin-exotic mesons with COMPASS}

\ShortTitle{Spin-exotic search with COMPASS}

\author{\speaker{Frank Nerling}\thanks{on behalf of the COMPASS collaboration}\\
        Physikalisches Institut, Albert-Ludwigs-Universit\"at Freiburg, 79104 Freiburg, Germany\\
        E-mail: \email{nerling@cern.ch}}

\abstract{The COMPASS fixed-target experiment at the CERN-SPS studies
 the structure and spectrum of hadrons. One important goal using
 hadron beams is the search for new states, in particular spin-exotic 
 mesons and glueballs. 
 As a first input to the puzzle, COMPASS observed a significant $J^{PC}$ spin-exotic signal 
 in the 2004 pilot run data (190\,GeV/$c$ $\pi^{-}$ beam, Pb target) in three charged pion 
 final states consistent with the disputed $\pi_1(1600)$. 
 We started our hadron spectroscopy programme in 2008 by collecting very 
 high statistics using a 190 GeV/$c$ negative pion beam scattered off a liquid hydrogen (proton) target.  
 The current status and new results from the 2008 data on the search for the $\pi_1(1600)$ resonance with exotic $J^{PC}=1^{-+}$ 
 quantum numbers obtained from partial-wave analyses of the $\rho\pi$ and $\eta'\pi$
 decay channels are presented.
}

\FullConference{The 2011 Europhysics Conference on High Energy Physics, EPS-HEP 2011,\\
		July 21-27, 2011\\
		Grenoble, Rh\^one-Alpes, France}

\begin{document}

\section{Introduction}
\vspace{-0.4cm}
The goal of the COMPASS experiment~\cite{compass-spectro} at the CERN-SPS is to obtain a better understanding of the 
structure and dynamics of hadrons, both aspects of non-perturbative Quantum Chromodynamics (QCD).
Given the accurate charged particle tracking as well as good coverage by electromagnetic calorimetry, the COMPASS 2008 hadron 
data provide an excellent opportunity for simultaneous observation of new states in different decay modes within 
the same experiment.

The existence of exotic states beyond the simple Constituent Quark Model (CQM), like so-called hybrids or glueballs, 
has been speculated about almost since the introduction of colour~\cite{Jaffe:1976,Barnes:1983}. 
Hybrid mesons are $q\bar{q}$ states with excited gluonic degree of freedom, whereas glueballs are purely gluonic states 
without valence quarks. They are both allowed within QCD (due to the self-coupling of gluons via the colour-charge), while 
they are forbidden within the simple CQM. 
Glue-ball candidates have been reported by the Crystal Barrel and the WA102 experiments, however, the mixing 
with ordinary isoscalar mesons complicates the interpretation. Several light hybrids on the other hand are predicted to 
have exotic $J^{PC}$ quantum numbers and are thus promising candidates in the search for resonances beyond the 
CQM. The hybrid candidate lowest in mass is predicted~\cite{Morningstar:2004} to have a mass between 1.3 and 2.2\,GeV/$c^2$ 
and spin-exotic quantum numbers $J^{PC}=1^{-+}$, not attainable by ordinary $q\bar{q}$ states.
In the light-quark sector, two $1^{-+}$ hybrid candidates have been experimentally observed in the past in different decay 
channels, the $\pi_1(1400)$ mainly seen in $\eta\pi$ decays, by e.g. E852~\cite{E852}, VES\cite{Beladidze:1993}, and Crystal 
Barrel~\cite{CB}, and the $\pi_1(1600)$, observed by both E852 and VES in the decay channels: $\rho\pi$~\cite{Adams:1998,Khokhlov:2000}, $\eta'\pi$~\cite{Beladidze:1993,Ivanov:2001}, $f_{1}\pi$~\cite{JHLee:1994,Kuhn:2004,Amelin:2005}, and $\omega\pi\pi$~\cite{Amelin:2005,Lu:2005}.
In $f_1\pi$, also a $\pi_1(2000)$ has been reported~\cite{Kuhn:2004}, however, this state still lacks any confirmation.
In particular the resonant nature of the $\rho\pi$ decay channel of the $\pi_1(1600)$ observed in $3\pi$ final states is highly disputed~\cite{Amelin:2005,Dzierba:2006}. COMPASS has started to shed new light on the puzzle of spin-exotics by the observation of an $1^{-+}$ signal in the 2004 data, consistent with the famous $\pi_1(1600)$. It shows clean phase motions with respect to other waves, confirming the resonance 
nature~\cite{Alekseev:2009a}.            
\begin{figure}[tp!]
  \begin{minipage}[h]{.32\textwidth}
    \begin{center}
\vspace{-0.7cm}
     \includegraphics[clip,trim= 3 4 22 5, width=1.0\linewidth, angle=90]{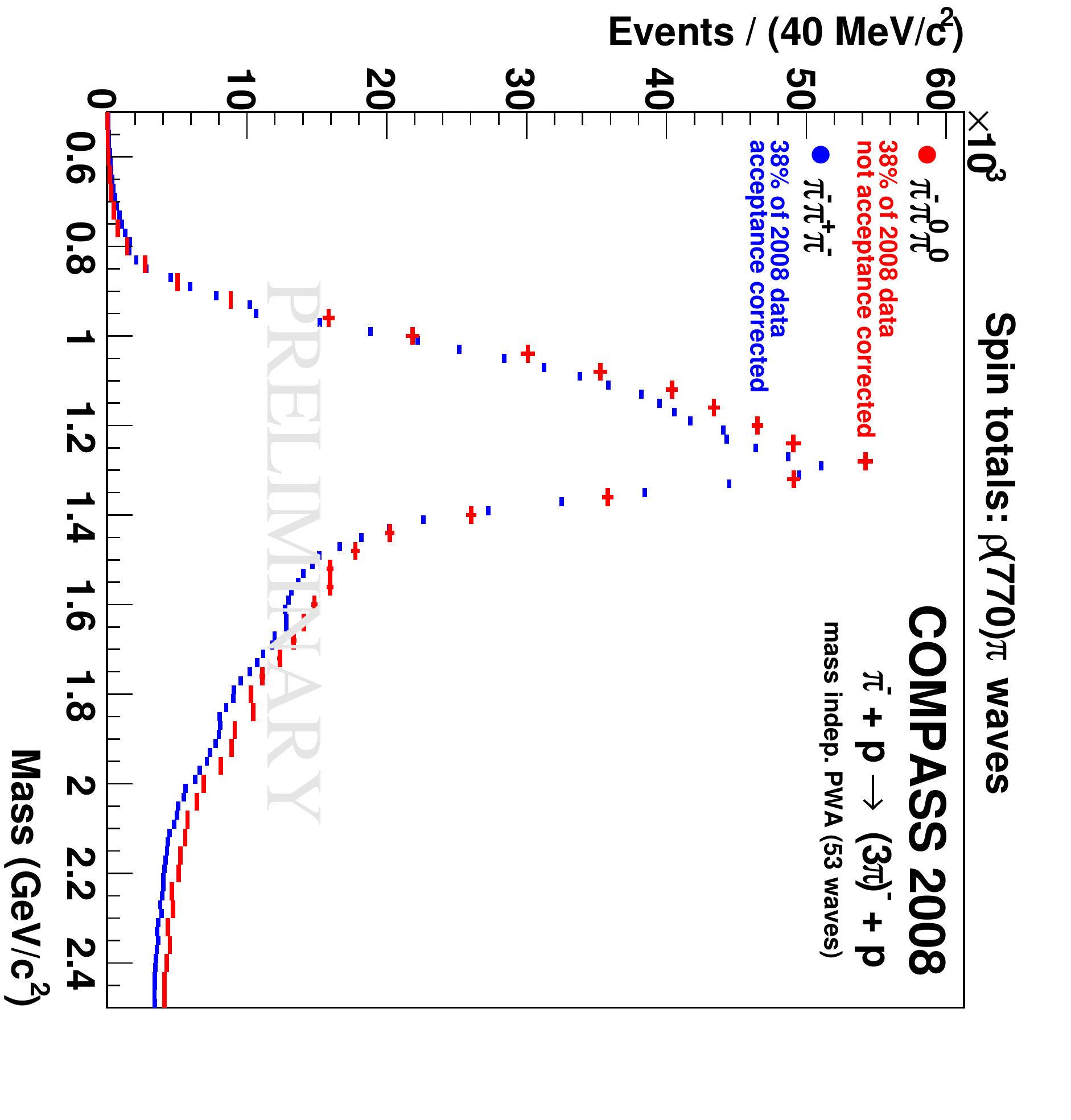}
    \end{center}
  \end{minipage}
  \hfill
  \begin{minipage}[h]{.32\textwidth}
    \begin{center}
\vspace{-0.7cm}
     \includegraphics[clip,trim= 3 4 22 5, width=1.0\linewidth, angle=90]{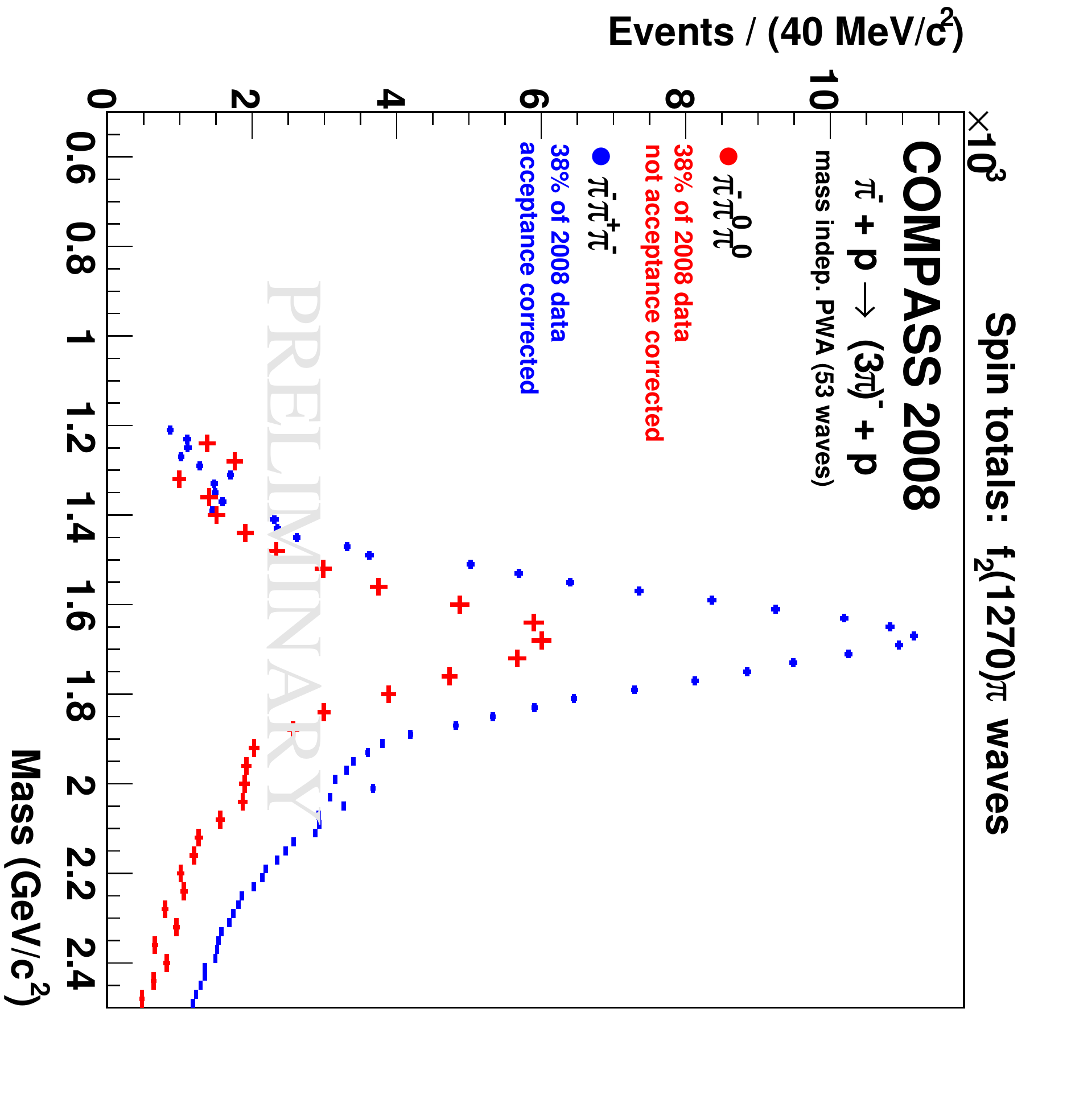}
    \end{center}
  \end{minipage}
  \hfill
  \begin{minipage}[h]{.32\textwidth}
    \begin{center}
\vspace{-0.7cm}
      \includegraphics[clip,trim= 5 20 23 10, width=1.0\linewidth, angle=90]{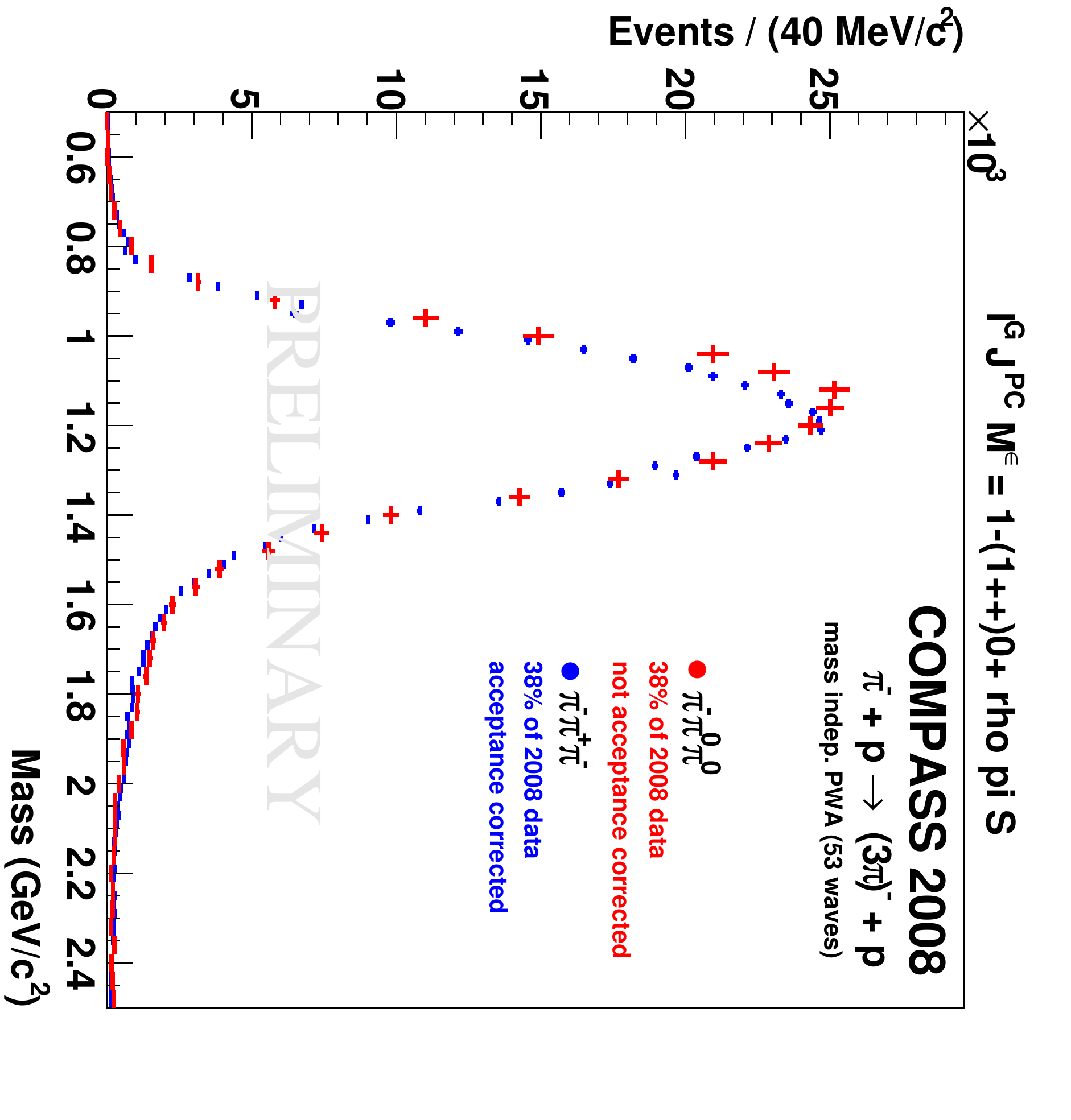}
    \end{center}
  \end{minipage}
      \vspace{-0.1cm}
     \caption{Intensity sums of all partial waves decaying into $\rho\pi$ (left) and $f_2\pi$ (centre) intermediate states included in the fits are in good agreement with expectations from isospin symmetry. The $a_2(1320)$ has been used for normalisation, see Fig.\,2 (top, left). The data follows isospin symmetry not only for main waves, exemplary shown is the $a_1(1260)$ (right), but also for small ones, 
cf. Fig.\,2, for discussion see text.}
       \label{fig:isospinSymmSpinTotals-53w}
\end{figure}
\section{New partial-wave analysis results on the search for the $\pi_1(1600)$}
\label{sec.PWAresults}
\begin{figure}[tp!]
  \begin{minipage}[h]{.32\textwidth}
    \begin{center}
         \vspace{-0.7cm}
     \includegraphics[clip,trim= 0 0 0 0, width=1.0\linewidth, angle=90]{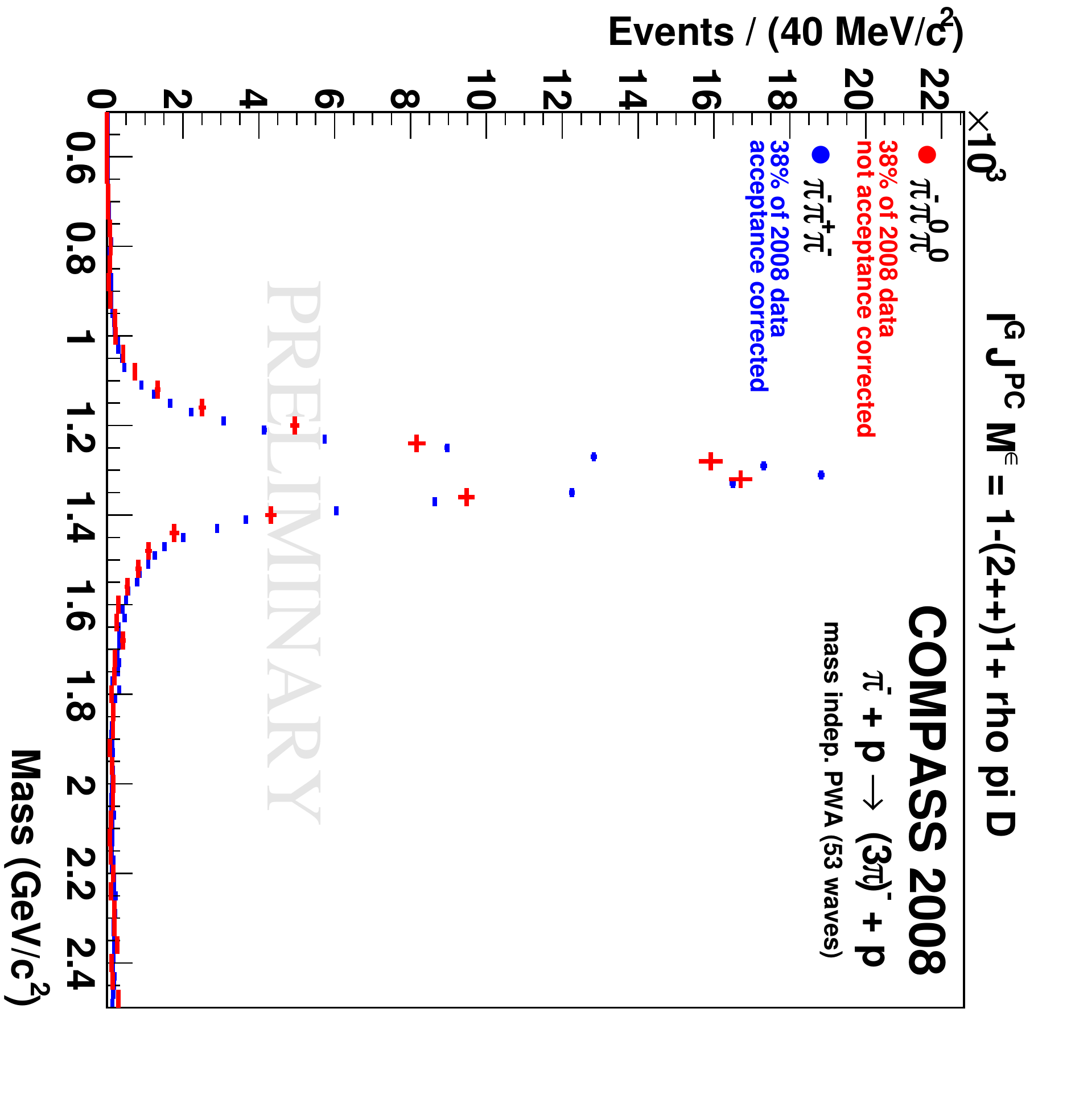}
    \end{center}
  \end{minipage}
  \hfill
  \begin{minipage}[h]{.32\textwidth}
    \begin{center}
      \vspace{-0.7cm}
     \includegraphics[clip,trim= 0 0 0 0, width=1.0\linewidth, angle=90]{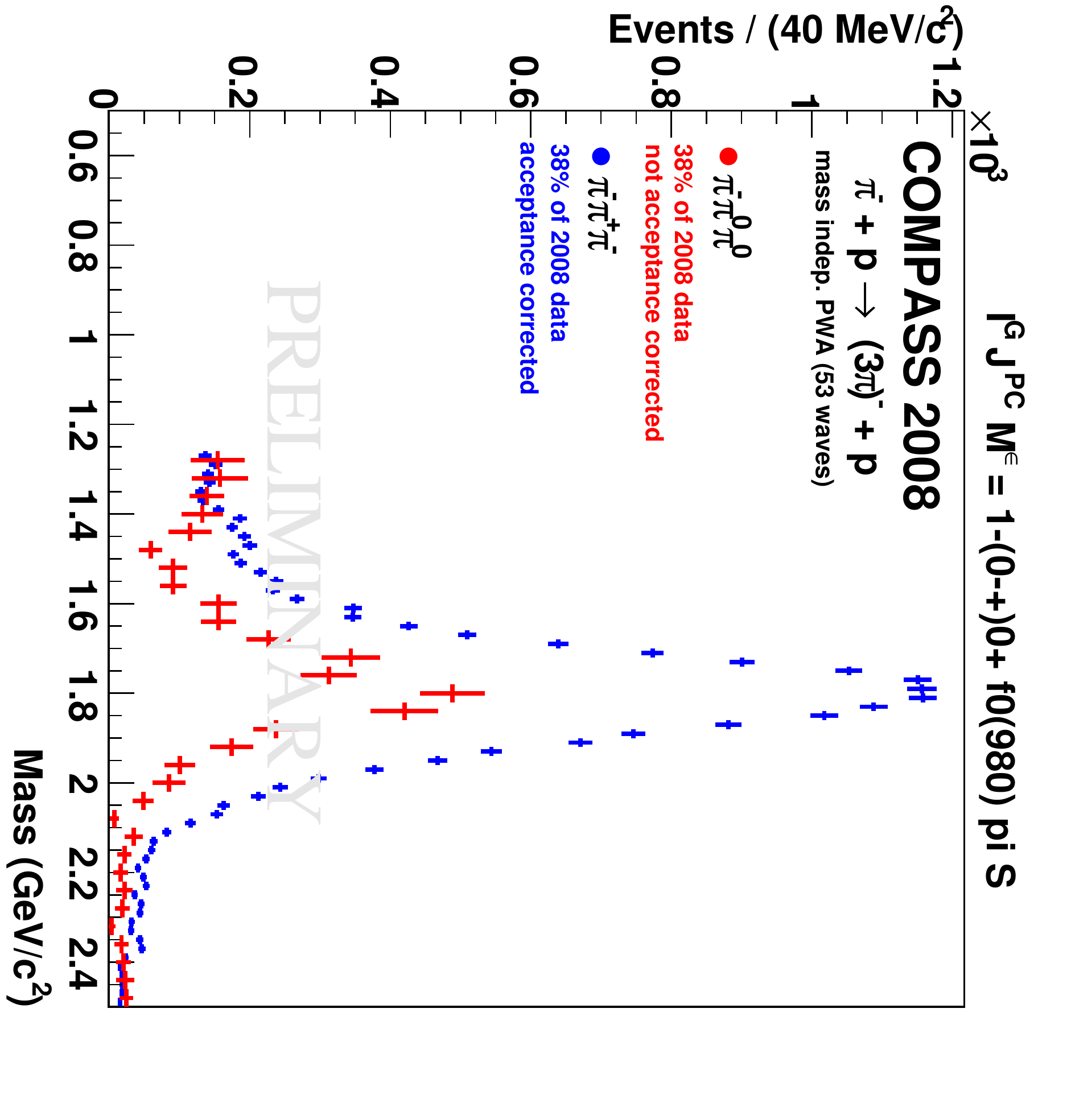}
    \end{center}
  \end{minipage}
  \hfill
  \begin{minipage}[h]{.32\textwidth}
    \begin{center}
      \vspace{-0.7cm}
     \includegraphics[clip,trim= 0 0 0 0, width=1.0\linewidth, angle=90]{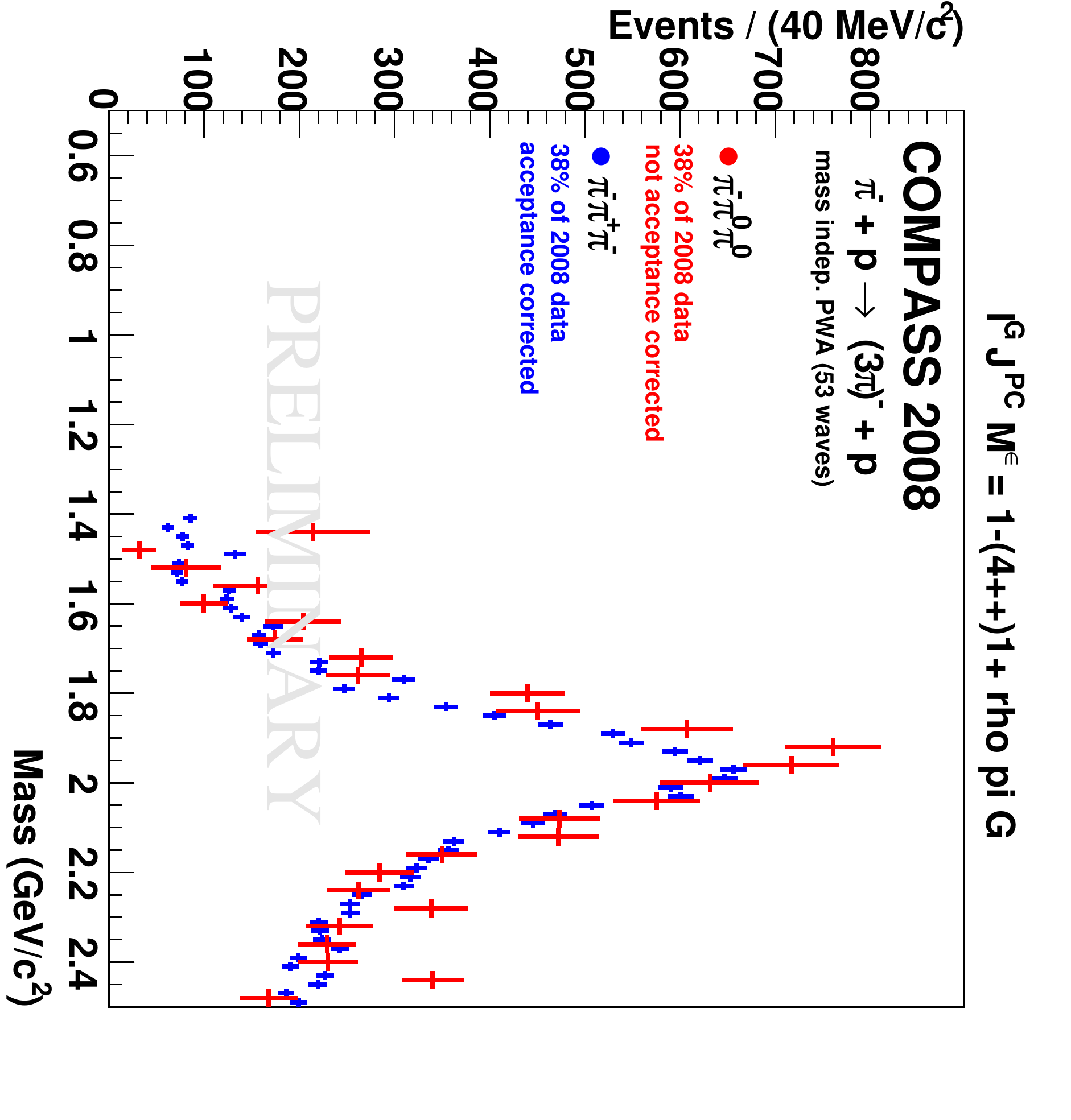}
    \end{center}
  \end{minipage}
  \begin{minipage}[h]{.32\textwidth}
    \begin{center}
     \vspace{-0.3cm}
     \includegraphics[clip,trim= 0 0 0 0, width=1.0\linewidth, angle=90]{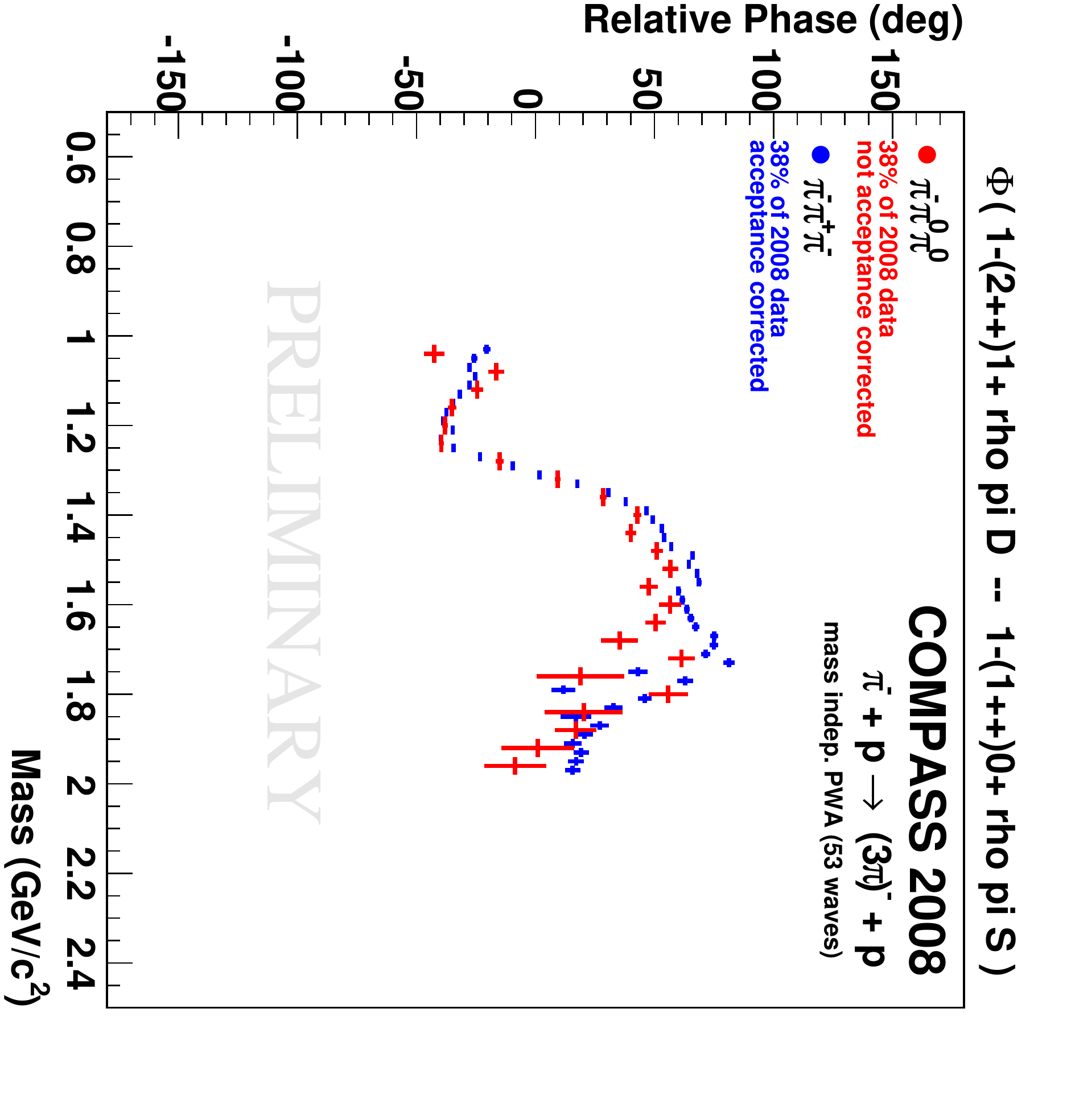}
    \end{center}
  \end{minipage}
  \hfill
  \begin{minipage}[h]{.32\textwidth}
    \begin{center}
      \vspace{-0.3cm}
     \includegraphics[clip,trim= 0 0 0 0, width=1.0\linewidth, angle=90]{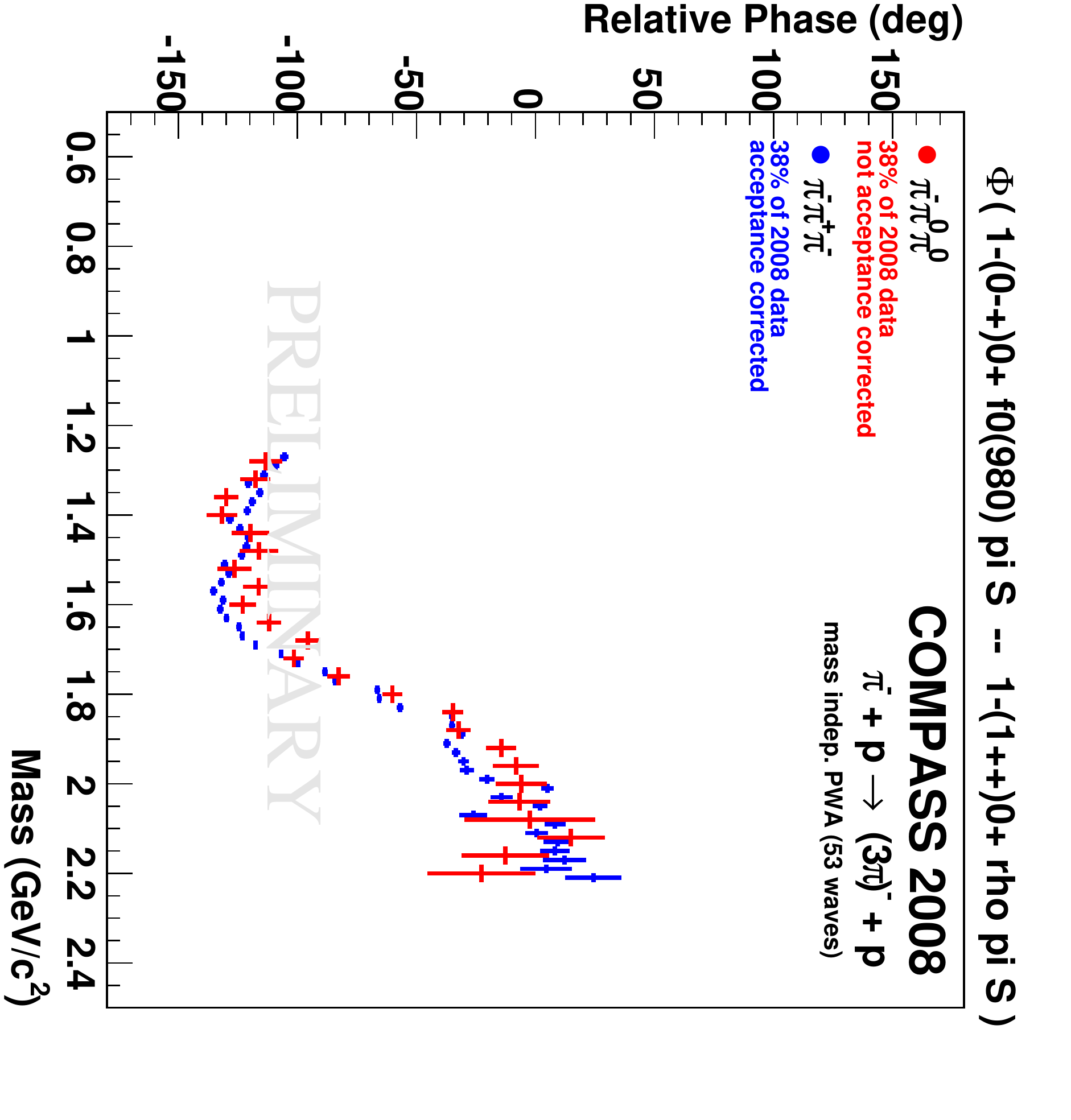}
    \end{center}
  \end{minipage}
  \hfill
  \begin{minipage}[h]{.32\textwidth}
    \begin{center}
      \vspace{-0.3cm}
      \includegraphics[clip,trim= 0 0 0 0, width=1.0\linewidth, angle=90]{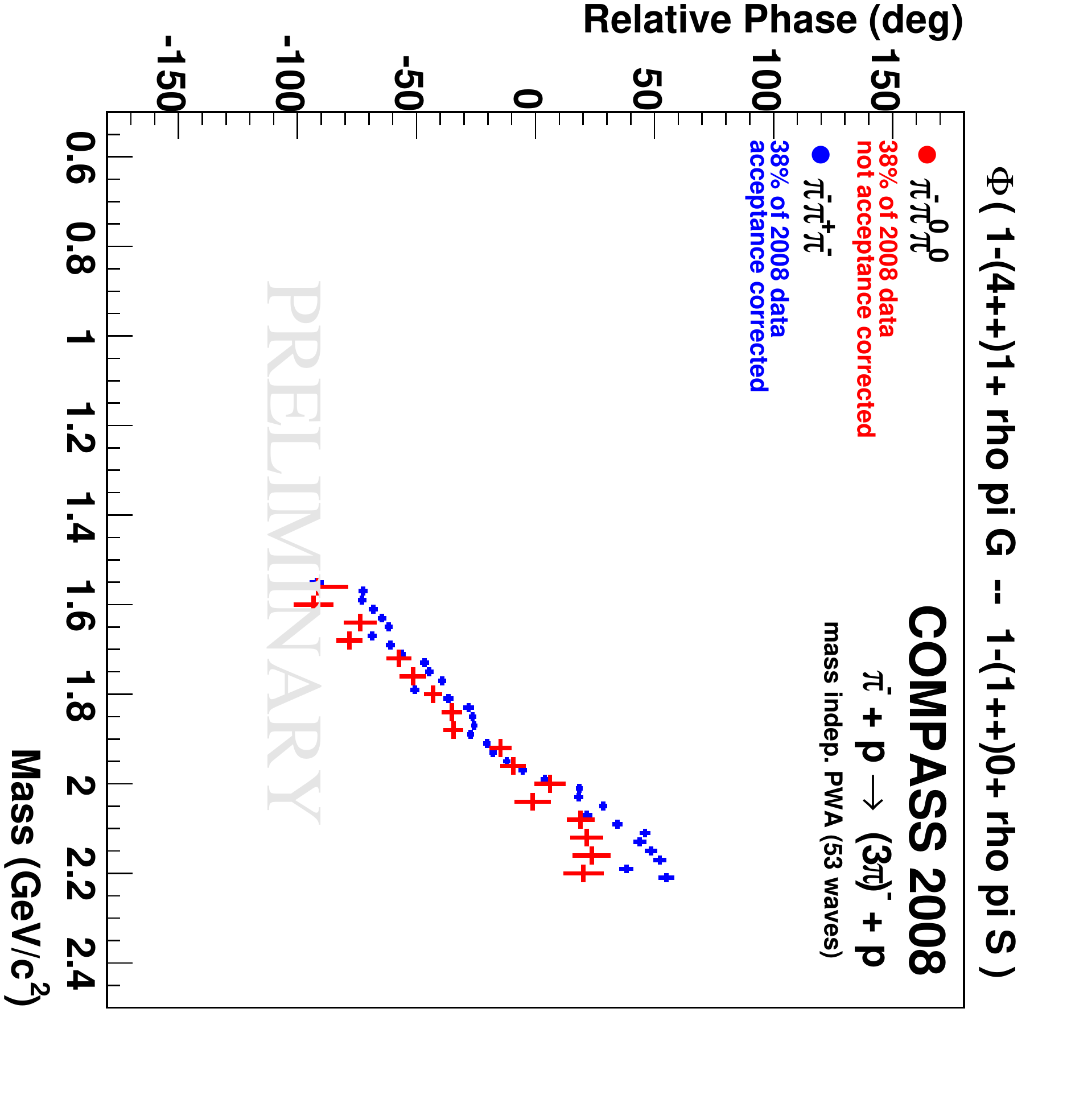}
    \end{center}
  \end{minipage}
  \begin{center}
      \vspace{-0.4cm}
     \caption{Intensities and relative phase differences $\Phi$ for main and small waves with respect to the prominent $a_1(1260)$. Exemplary shown are, from left to right, the $a_2(1320) \rightarrow \rho\pi$, the $\pi(1800) \rightarrow f_0(980)\pi$, and the $a_4(2040)\rightarrow \rho\pi$, respectively. The well-known $a_2(1320)$ resonance as well as the smaller, less prominent ones show a clean, rapid phase motion relatively to the 
$a_1(1260)$. Not only the intensities but also the phases are consistent for both, neutral ($\pi^{-}\pi^{0}\pi^{0}$) and charged ($\pi^{-}\pi^{+}\pi^{-}$) decay modes, for discussion see text.} 
    \label{fig:phases_a1_a2__a1_pi2-53w}
  \end{center}
  \vspace{-0.7cm}
\end{figure}

\vspace{-0.4cm}
COMPASS has access to all decay channels light spin-exotic mesons have been reported in so far, for a recent review 
see~\cite{MeyerHaarlem:2010}.
First PWA analyses focus on the $(3\pi)^{-}$ and $\eta'\pi^{-}$ final states, studying the existence 
of the contribution of the spin-exotic $1^{-+}$ wave in the 2008 data. 
\paragraph{Diffraction of $\pi^{-}$ into $(3\pi)^{-}$ final states -- neutral and charged mode ($\rho\pi$ decay channel)}
The new mass-independent PWA results for neutral and charged mode data presented in this paper, are
normalised using the $a_2(1320)$ as a standard candle as shown in Fig.\,\ref{fig:phases_a1_a2__a1_pi2-53w} (top, left)
in order to compensate for the different detection efficiencies. This makes the fitted intensities for individual 
partial waves comparable between neutral and charged decay modes. A detailed description of the applied PWA method can be found 
in~\cite{nerling:2009,haas:2011} and references therein.

Due to the Clebsch-Gordan coefficients determining the different isospin coupling for the different underlying 
isobar structure, i.e. decays into an isovector versus an isoscalar, we expect for isospin 1 resonances decaying 
to $\rho\pi$ similar intensities for the neutral and charged mode data, whereas decays into $f_2\pi$ should show 
a suppression factor of two. Even though the neutral mode data have not yet been corrected for acceptance (similarly 
flat for the charged mode as already for the 2004 data), the data is in good agreement with these expectations from 
simple isospin coupling considerations. 
This holds to a large extent throughout the whole wave-set as shown by Fig.\,\ref{fig:isospinSymmSpinTotals-53w}, 
depicting the intensity sums of all $\rho\pi$ (left) and $f_2\pi$ partial waves (centre), respectively.
A few examples for fitted intensities of individual main and small waves are given in Fig.\,\ref{fig:isospinSymmSpinTotals-53w} 
(right), showing the $a_1(1260)$ decaying into $\rho\pi$ observed with same width and intensity for both decay modes, and 
Fig.\,\ref{fig:phases_a1_a2__a1_pi2-53w}, respectively, where also the relative phases with respect to the $a_1(1260)$ are shown.
For the $\pi(1800)$ decay into $f_0(980)\pi$, we rather find about 45\,\% instead of 50\,\% for the intensity of the neutral as 
compared to the charged mode data, which is indeed expected taking into account effects from Bose-Symmetrisation, for a detailed 
discussion see~\cite{nerling:2011}.

\begin{figure}[tp!]
  \begin{minipage}[h]{.49\textwidth}
    \begin{center}
\vspace{-0.7cm}
     \includegraphics[clip,trim= 0 0 0 0, width=1.0\linewidth]{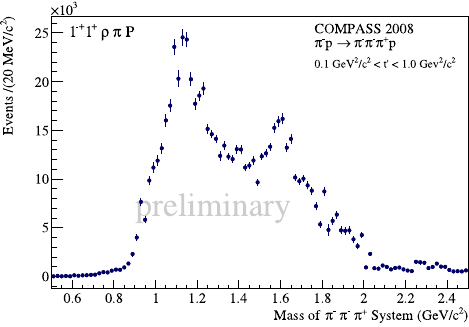}
    \end{center}
  \end{minipage}
  \hfill
  \begin{minipage}[h]{.49\textwidth}
    \begin{center}
\vspace{-0.7cm}
     \includegraphics[clip,trim= 0 0 0 0, width=1.0\linewidth, angle=0]{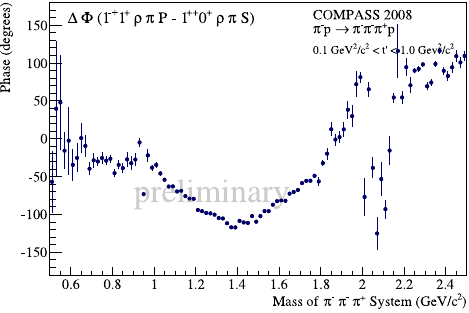}
    \end{center}
  \end{minipage}
      \vspace{-0.2cm}
     \caption{Intensity of the spin-exotic $1^{-+}$ wave (left) and the phase difference with respect to the $a_1(1260)$.}
       \label{fig:exoticWaveCharged-53w}
  \vspace{-0.3cm}
\end{figure}

The fitted intensities of the spin-exotic $1^{-+}$ wave and the relative phase again with respect to the 
$a_1(1260)$ resonance are shown for the charged mode (Fig.\,\ref{fig:exoticWaveCharged-53w}). 
This wave shows a structure at about 1.6\,GeV/$c^{2}$. A raising phase is observed in the corresponding mass region, consistent 
with the structure in the exotic wave at 1.6\,GeV/$c^{2}$ having resonant nature and resonating against the tail of the $a_1(1260)$. 
This is consistent with the result of the PWA of the 2004 pilot run data taken with a Pb target~\cite{Alekseev:2009a}. The bump at 
around 1.2\,GeV/$c^{2}$ needs further to be understood, it appears unstable with respect to changes in the PWA model, and is 
still under investigation~\cite{haas:2011}. 

\begin{figure}[tp!]
\vspace{-0.7cm}
  \begin{center}
     \includegraphics[clip, trim= 75 70 75 210,width=1.0\linewidth]{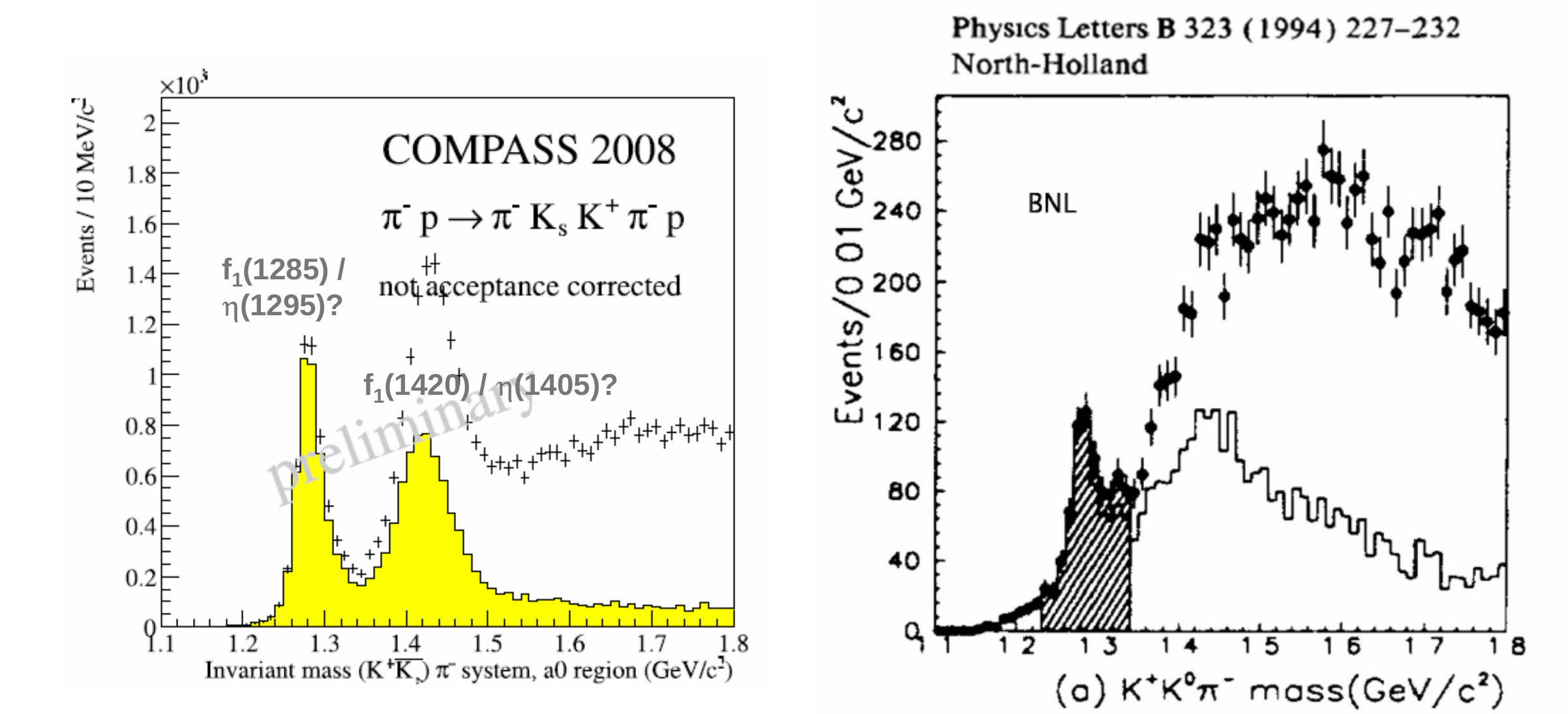}
    \end{center}
\vspace{-0.7cm}
      \caption{The $(K\bar{K}\pi)^{0}$ sub-mass system shows clean $f_1(1285)$ and $f_1(1420)$ peaks 
before (dots) and after (line) an additional restriction of the $(K\bar{K})$ mass to the $a_0(980)$ region. 
Comparing the similar plot obtained at BNL~\cite{JHLee:1994} (right), the COMPASS statistics exceeds 
the one analysed by E852 by a factor of 10, even a factor of 20, taking into account also the 2009 data taken 
with $\pi^{-}$ beam. Not only the observed $f_1(1285)$ but also the $f_1(1420)$ are nearly 
background free as compared to the published result by BNL/E852~\cite{JHLee:1994}.}
       \label{fig:KKpipi} 
\vspace{-0.5cm}
\end{figure}

\paragraph{Diffraction of $\pi^{-}$ into $\pi^{-}\pi^{+}\pi^{-}\eta$ final states ($\eta'\pi$ decay channel)}
Reconstructing $\pi^{-}\pi^{+}\pi^{-}\eta$ final states, the diffractively produced $\pi^{-}\eta'$ system can be 
analysed using PWA. 
\begin{figure}[bp!]
  \begin{center}
\vspace{-0.7cm}
     \includegraphics[clip, trim= 65 20 55 80,width=1.0\linewidth]{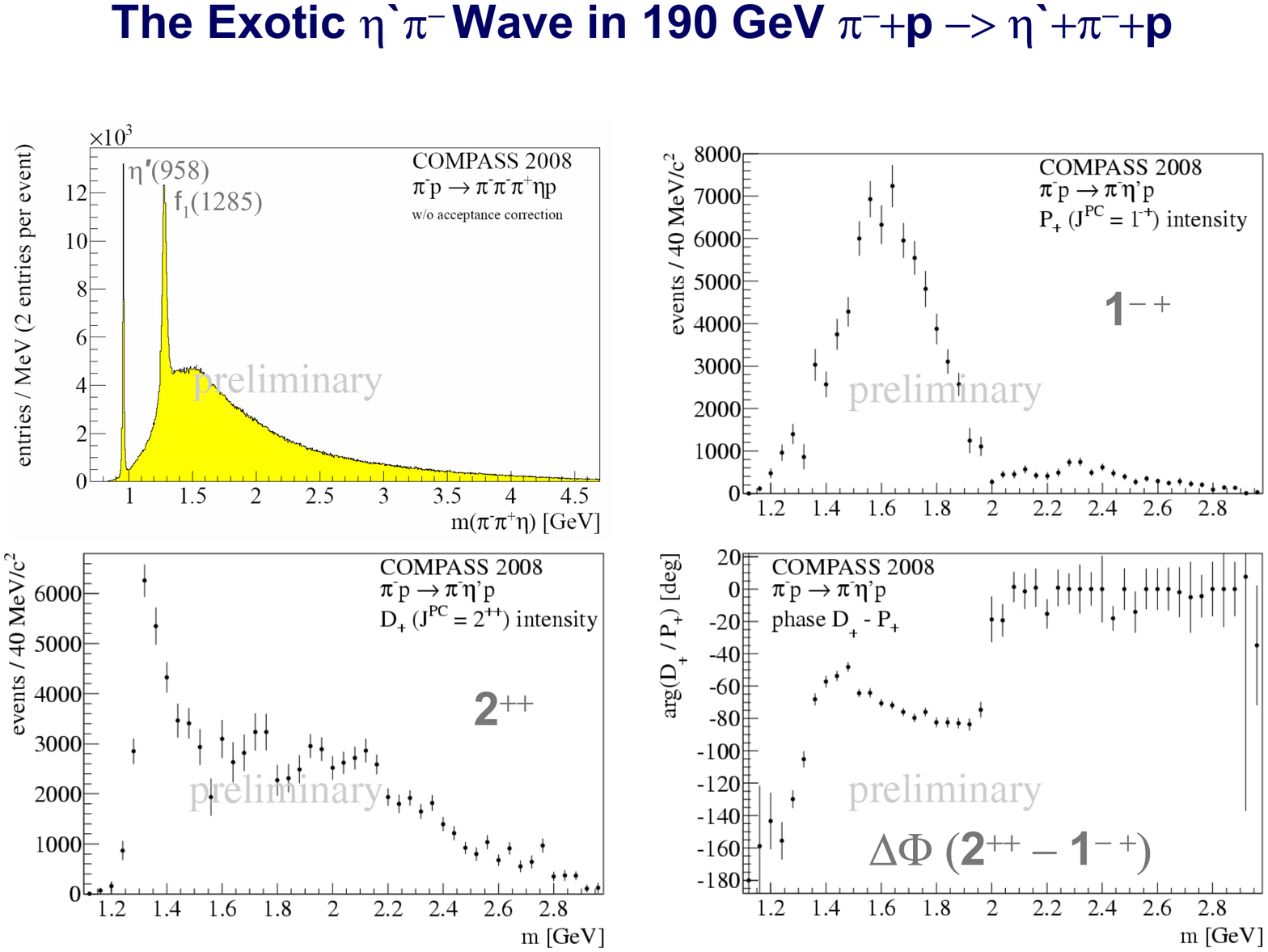}
    \end{center}
\vspace{-0.7cm}
      \caption{Top: Invariant sub-mass spectrum showing $\eta'(958)$ and $f_1(1285)$ peaks (left), and the mass-independently fitted intensities of the spin-exotic $1^{-+}$ ($P_+$) wave (right). Bottom: Fitted intensities of the $2^{++}$ ($D_+$) wave showing the $a_2(1320)$ (left) and the relative phase $D_+-P_+$ (right).}
       \label{fig:eta_prime_pi} 
\vspace{-0.7cm}
\end{figure}
The involved decay chain is here $\eta' \rightarrow \pi^{+}\pi^{-}\eta, \eta \rightarrow \gamma\gamma$, 
and the resulting sub-mass spectrum, Fig.\,\ref{fig:eta_prime_pi} (top, left), exhibits clean, narrow $\eta'(958)$ 
and $f_1(1285)$ peaks. The mass-independent PWA follows previous analyses, $P_{+}$ and $D_{+}$ 
waves are included as well as a $G_{+}$ wave in addition~\cite{tobi:2011}.   
A clear signal of the $a_2(1320)$ ($D_+$) appears, and the most intense wave is the spin-exotic ($P_+$) wave, showing a broad structure at around 1.6\,GeV/$c^2$ consistent with other experiments. More systematic studies are, however, needed before strong conclusions can be drawn, the sharp jump in the phase difference ($D_+ - P_+$) at about 2\,GeV/$c^2$ has e.g. further to be understood. 
\paragraph{Diffraction of $\pi^{-}$ into $K^{0}_{s}K^{\pm}\pi^{\mp}\pi^{-}$ final states ($f_1\pi$ decay channel) -- Outlook}
The $(K\bar{K}\pi)^{0}$ subsystem shown in Fig.\,\ref{fig:KKpipi} (exemplary for one case) is of particular interest, 
as spin-exotic $1^{-+}$ resonances were reported in the $f_1\pi$ decay channel. The COMPASS data feature clean $f_1(1285)$ and 
$f_1(1420)$ peaks (Fig.\,\ref{fig:KKpipi}). Even though an $\eta$ contribution cannot be excluded, a first mass-independent PWA 
indicate contributions from $\eta(1405)$ and $\eta(1295)$ to be minor, consistent with the observation by 
E852~\cite{Kuhn:2004}, for further details, see~\cite{bernhard_nerling:2011}. 
The complementary possibility for studying the $f_1\pi$ decay channel are $\eta 3\pi$ final states, also analysed by 
E852. This channel is feasible (Fig.\,\ref{fig:eta_prime_pi}, top/left) and will be analysed by COMPASS as well. 
%
\vspace{-0.5cm}
\section{Conclusions \& summary}
\vspace{-0.4cm}
The high statistics hadron data will allow COMPASS to contribute solving the puzzle of light 
spin-exotic mesons. The new results presented on the $\rho\pi$ decay channel in both, the neutral and charged decay 
modes of the $(3\pi)^{-}$ system, appear very consistent and solid not only for main but also for small waves.
There is presently no contradiction between both analyses results. The result from the 2004 pilot run data seems 
on the way to be confirmed, more systematic studies are needed before doing the mass-dependent PWA analyses. 
In-line with other experiments, a huge contribution of the spin-exotic $1^{-+}$ wave in the $\eta'\pi$ 
decay channel is confirmed in the COMPASS data, even though more systematic studies are needed before conclusions 
on the resonant nature can be drawn. 
The feasibility for spin-exotic search in the $f_1\pi$ decay channel in diffractively produced $K\bar{K}\pi\pi$ final states 
has been shown, not only for $f_1(1285)\pi$ but also for a first study of the $f_1(1420)\pi$ system, never done before.     
Apart of the results discussed in this paper, COMPASS has all important decay channels for spin-exotic search at sufficient statistics on 
tape, and the recent results presented here confirm the excellent potential to conclude on the existence of the 
highly disputed spin-exotic $\pi_1(1600)$ resonance and others. 
   
\paragraph{Acknowledgements}
This work is supported by the BMBF (Germany), in particular via the\\ ``Nutzungsinitiative CERN''.


\begin{thebibliography}{99}
\bibitem{compass-spectro} P.~Abbon {\it et al.}, COMPASS collaboration, {\it Nucl. Instr. Meth. A} {\bf 68} (2007) 455.
\bibitem{Jaffe:1976} R.~Jaffe and K.~Johnsons, {\it Phys. Lett. B} {\bf 60} (1976) {201}.
\bibitem{Barnes:1983} T.~Barnes {\it et al.}, {\it Nucl. Phys. B} {\bf 224} (1983) {241}.
\bibitem{Morningstar:2004} K.J.~Juge, J.~Kuti, C.~Morningstar, {\it AIP Conf. Proc.} {\bf 688} (2004) {193}.
\bibitem{E852} D.~R.~Thomson {\it et al.}, {\it Phys. Rev. Lett.} {\bf 79} (1997) 1630.
\bibitem{Beladidze:1993} G.~M.~Beladidze {\it et al.}, {\it Phys. Lett. B} {\bf 313} (1993) 276.
\bibitem{CB} A.~Abele {\it et al.}, {\it Phys. Lett. B} {\bf 423} (1998) 175.
\bibitem{Adams:1998} G.~S.~Adams {\it et al.}, {\it Phys. Rev. Lett.} {\bf 81}, (1998) 5760.
\bibitem{Khokhlov:2000} Y.~Khokhlov, {\it Nucl. Phys. A} {\bf 663} (2000) 596.
\bibitem{Ivanov:2001} E.~I.~Ivanov {\it et al.}, {\it Phys. Rev Lett.} {\bf 86} (2001) 3977.
\bibitem{JHLee:1994} J.H.~Lee {\it et al.}, {\it Phys. Lett. B} {\bf 323} (1994) {227}.
\bibitem{Kuhn:2004} J.~Kuhn {\it et al.}, {\it Phys. Lett. B} {\bf 595} (2004) 109.
\bibitem{Amelin:2005} D.~V.~Amelin {\it et al.}, {\it Phys. Atom. Nucl.} {\bf 68} (2005) 359.
\bibitem{Lu:2005} M.~Lu {\it et al.}, {\it Phys. Rev. Lett.} {\bf 94} ({2005}) {032002}.
\bibitem{Dzierba:2006} A.R.~Dzierba {\it et al.}, {\it Phys. Rev. D} {\bf 73} (2006) {072001}.
\bibitem{Alekseev:2009a} M.~Alekseev {\it et al.}, COMPASS collaboration,   
{\it Phys. Rev. Lett}, {\bf 104} (2010) {241803}.
\bibitem{MeyerHaarlem:2010} C.A.~Meyer and Y.Van Haarlem, {\it Phys. Rev. C} {\bf 82} (2010) {025208}; arXiv:1004.5516v2 [nucl-ex].
\bibitem{nerling:2009} F.~Nerling, {\it AIP Conf. Proc.} {\bf 1257} (2010) 286; arXiv:1007.2951 [hep-ex].
\bibitem{haas:2011} F.~Haas, {\it Conf. Proc. Hadron2011}, Munich, Germany (2011).
\bibitem{nerling:2011} F.~Nerling, {\it Conf. Proc. Hadron2011}, Munich, Germany (2011); arXiv:1108.5969 [hep-ex].
\bibitem{tobi:2011} T.~Schl\"uter, {\it Conf. Proc. Hadron2011}, Munich, Germany (2011); arXiv:1108.6191 [hep-ex].
\bibitem{bernhard_nerling:2011} J.~Bernhard and F.~Nerling, {\it Conf. Proc. Hadron2011}, Munich, Germany (2011); arXiv:1109.0219~[hep-ex].
\end{thebibliography}
\end{document}